\documentstyle{grf2004}

\begin{document} 

\topmargin=1cm

\begin{frontmatter}

\title{Charge conjugation and Lense-Thirring Effect}
\vspace{-36pt} 
\title{\textemdash~A new Asymmetry ~\textemdash}

\author{D. V. Ahluwalia-Khalilova}

\address{
Ashram for the Studies of the Glass Bead Game (ASGBG)\\
Center for Studies of Physical, Mathematical and 
Biological Structure of Universe \\
Department of Mathematics, Ap. Postal C-600\\
University of Zacatecas (UAZ), Zacatecas, ZAC 98062, Mexico\\
E-mail: dva-k@heritage.reduaz.mx}

\begin{abstract}\\

\noindent
This essay presents a new asymmetry that arises from the interplay of charge 
conjugation and Lense-Thirring effect. When applied to Majorana neutrinos, the 
effects predicts  $\nu_e \rightleftharpoons \overline{\nu}_e$ oscillations in 
gravitational environments with rotating sources. Parameters associated with 
astrophysical environments indicate that the presented effect is presently 
unobservable for solar neutrinos. But, it will play an important role in 
supernova explosions, and carries relevance for the observed matter-antimatter 
asymmetry in the universe.

\end{abstract}

\end{frontmatter}

%....................................... Various boldface symbols

\def\s{\mbox{\boldmath$\displaystyle\mathbf{\sigma}$}}
\def\J{\mbox{\boldmath$\displaystyle\mathbf{J}$}}
\def\K{\mbox{\boldmath$\displaystyle\mathbf{K}$}}
\def\A{\mbox{\boldmath$\displaystyle\mathbf{A}$}}
\def\B{\mbox{\boldmath$\displaystyle\mathbf{B}$}}

\def\P{\mbox{\boldmath$\displaystyle\mathbf{P}$}}
\def\p{\mbox{\boldmath$\displaystyle\mathbf{p}$}}
\def\hp{\mbox{\boldmath$\displaystyle\mathbf{\widehat{\p}}$}}
\def\x{\mbox{\boldmath$\displaystyle\mathbf{x}$}}
\def\0{\mbox{\boldmath$\displaystyle\mathbf{0}$}}
\def\bv{\mbox{\boldmath$\displaystyle\mathbf{\varphi}$}}
\def\hbv{\mbox{\boldmath$\displaystyle\mathbf{\widehat\varphi}$}}

\def\bn{\mbox{\boldmath$\displaystyle\mathbf{\nabla}$}}

\def\bl{\mbox{\boldmath$\displaystyle\mathbf{\lambda}$}}
\def\bl{\mbox{\boldmath$\displaystyle\mathbf{\lambda}$}}
\def\bfhh{\mbox{\boldmath$\displaystyle\mathbf{(1/2,0)\oplus(0,1/2)}\,\,$}}

\def\mn{\mbox{\boldmath$\displaystyle\mathbf{\nu}$}}
\def\amn{\mbox{\boldmath$\displaystyle\mathbf{\overline{\nu}}$}}

\def\mne{\mbox{\boldmath$\displaystyle\mathbf{\nu_e}$}}
\def\amne{\mbox{\boldmath$\displaystyle\mathbf{\overline{\nu}_e}$}}
\def\rlh{\mbox{\boldmath$\displaystyle\mathbf{\rightleftharpoons}$}}

\def\wm{\mbox{\boldmath$\displaystyle\mathbf{W^-}$}}
\def\hh{\mbox{\boldmath$\displaystyle\mathbf{(1/2,1/2)}$}}
\def\h00h{\mbox{\boldmath$\displaystyle\mathbf{(1/2,0)\oplus(0,1/2)}$}}
\def\znbb{\mbox{\boldmath$\displaystyle\mathbf{0\nu \beta\beta}$}}

%.......................

\def\rb{\kappa^{\left(\frac{1}{2},0\right)}}
\def\lb{\kappa^{\left(0,\frac{1}{2}\right)}}

%........................ For Eqs. in Sec.3
\def\rf{\frac{m^2+p^2-E^2}{(m-p+E)(m+p+E)\sqrt{2 m (m+E)}}}
\def\aa{m+E+p\cos(\theta)}
\def\bb{m+E-p\cos(\theta)}
\def\ab{p\, e^{-i\phi}  \sin(\theta)}
\def\ba{p \,e^{i\phi}  \sin(\theta)}
%........................

\def\beq{\begin{eqnarray}}
\def\eeq{\end{eqnarray}}

\def\ua{\{-,+\}}
\def\da{\{+,-\}}
\def\defn{\buildrel \rm def \over =}

%........................................................

\textit{Introduction}.\textemdash~Consider a rotating matter sphere charged with electrons.
Then, consider a similarly rotating antimatter sphere charged with
positrons. Formally, the latter is obtained from the former,
and vice versa, by the
action of an appropriate set of charge conjugation operators
associated with matter and gauge fields.
It is a purely general-relativistic  prediction
that in this transformation from matter to antimatter, 
the gravitational Lense-Thirring moment \cite{Lense:1918lte}
\textemdash~
or, the so called gravitomagnetic moment \textemdash~ 
does not alter its direction, while
the associated magnetic moments flip their directions. 
If for the matter sphere
the magnetic and  Lense-Thirring moments are chosen to 
be \textit{parallel}, for the antimatter sphere they become 
\textit{anti-parallel}. This suggests that 
gravity may carry an intrinsic charge conjugation asymmetry. 
This is a brief summary \textemdash~missing operational
 considerations 
\textemdash~
 of an argument which I presented in late 1983, or perhaps
early 1984, in a conversation with Kip Thorne at Texas 
A\&M University.

Now, independently, work of Singh and collaborators hints 
at a similar asymmetry 
\cite{Mukhopadhyay:2003ij,Lambiase:2003fd}. 
If these gravitationally-induced asymmetries were to be true
it would have profound consequences for theoretical physics as well as 
astrophysical and cosmological processes. I now pursue, and establish, 
one of these asymmetries in a manner which shall reduce it to its bare 
essentials without any reliance on a particular theory of quantum gravity. 

\textit{Establishing the thesis}. \textemdash~ Even though it is not a conventional wisdom for most 
general relativists, the primitive notion of test particles 
derives its definition from,
(a) the Casimir invariants of the Poincare group, and 
(b) local gauge symmetries of the associated wave 
equations.\footnote{In the conventional general relativistic framework it is implicitly assumed
that the nature of test particle is hardly of significance as long as it satisfies some
very primitive requirements of size, etc. The only exception that we know are
inclusion of spin/internal-structure effects on geodesic deviations where the
work of Anandan \textit{et al.} and Mohseni provide latest developments on the subject \cite{Anandan:2003ev,Mohseni:2004br}.}
The interplay of these two, for instance, gives
the possibility to have  massive spin\textendash$1/2$ test particles.
There are two type of spin\textendash$1/2$ particles:
Those which are eigenstates of the charge operator, $\mathcal{Q}$,
and those which are eigenstates of the charge conjugation 
operator, $\mathcal{C}$. The former are governed by the Dirac equation,
while 
the latter eigenspinors 
\textemdash~ note, $\left[\mathcal{C},\,\mathcal{Q}\right]\ne 0$, 
in general 
\textemdash~ 
we must construct \textit{ab initio}.

The eigenspinors of the $\mathcal{C}$  have the property 
that \cite{ahluwalia:abr}:
$
\mathcal{C}  \,\lambda(\p) = \pm\, \lambda(\p)
$.
Those corresponding to the \textit{plus} sign, 
we call self conjugate, are given by:
\beq
\lambda^S(\p) =
\left(
\begin{array}{c}
\sigma_2 \phi^\ast_L(\p)\\
\phi_L(\p)
\end{array}
\right)\,,
\eeq
 where $\phi_L(\p)$ is a massive, left-handed, Weyl spinor.
These are the standard Majorana spinors found in textbooks 
\cite{Ramond:1989prb}. Those corresponding to the \textit{minus} 
sign, we call  anti-self conjugate, and these read: 
\beq
\lambda^A(\p) =
\left(
\begin{array}{c}
- \sigma_2 \phi^\ast_L(\p)\\
\phi_L(\p)
\end{array}
\right)\,.
\eeq
These are new. The two sets differ by a physically important relative phase
between the right-handed transforming components, $\sigma_2 \phi^\ast_L(\p)$,
and the left-handed transforming components, $\phi_L(\p)$.
The $\lambda^A(\p)$ must complement $\lambda^S(\p)$ in order to have
a mathematically
complete representation space. Without $\lambda^A(\p)$ no
correct physical interpretation exists for the massive case.\footnote{See, Appendix for more details and 
brief interpretational remarks.}

To arrive at the asymmetry induced by gravitational Lense-Thirring field, 
let  
$\phi_L(\p)$ be eigenstates of the helicity operator,
$h\defn \s\cdot\hp$.
That is, 
\beq
h \;\phi_L^\pm (\p) = \pm \;\phi_L^\pm (\p)\,.
\eeq
With these definitions, it is now a simple mathematical exercise to
show that, 
\beq
h \;\sigma_2 {\phi_L^\ast}^\pm (\p) = \mp \;\sigma_2 {\phi_L^\ast}^\pm 
(\p)\,.\label{result}
\eeq

Stated in words:  \textit{The left- and right- transforming 
spinor components
of $\lambda(\p)$, irrespective of self/antiself conjugacy,
carry \underline{opposite} helicities. 
An eigenspinor of $\mathcal{C}$ is a dual helicity object.}

It is this circumstance which profoundly distinguishes
the eigenspinors of $\mathcal{C}$ from those of $\mathcal{Q}$.
The
eigenspinors of the latter are definite \textemdash~ 
i.e., single,  as opposed to
dual \textemdash~ helicity objects.
There are four eigenspinors of $\mathcal{C}$. These can be enumerated
as: 
\beq
\lambda^S_{\{-+\}}(\p),\; \lambda^S_{\{+-\}}(\p),\; \lambda^A_{\{-+\}}(\p),
 \;\mbox{and}\; \lambda^A_{\{+-\}}(\p)\,,
\eeq
where the first symbol in the subscript refers
to the helicity of the 
right-transforming spinor component; while the second  refers
to the helicity of the 
left-transforming spinor component. Now if we introduce, say the state associated
with $ \lambda^S_{\{-+\}}(\p)$ in the gravitational environment of a 
(roughly) spherically symmetric astrophysical source of mass, $M$, and angular
frequency $\omega$ \textemdash~ which contributes Lense-Thirring component to the 
gravitational field \textemdash~ then there are two physically relevant phase factors 
that such a state picks up:
\begin{enumerate}
\item[$\alpha$.]
An overall gravitationally-induced phase 
which depends on source mass, $M$, and energy, $E$, of the test particle.
 
\item[$\beta$.]
A gravitationally-induced \textit{relative} phase 
between the right- and left- transforming
components of the test particle.
\end{enumerate} 
The first of these enumerated phases has no effect for a single mass
eigenstate. However, it is precisely this phase when considered in the context of
neutrinos \textemdash~ which the solar, atmospheric, and other neutrino oscillation data 
\cite{Sanchez:2003rb,Bellerive:2004dm,Kato:2003qx,Smy:2002rz,Habig:2001ej} establishes to be 
linear superposition of three different mass eigenstates in 
a leptonic-flavor dependent manner \textemdash~ that flavor oscillation clocks gravitationally 
redshift as required by general relativity \cite{Ahluwalia:1996ev,Ahluwalia:1998jx,Konno:kq,Wudka:2000rf,Adak:2000tp,Crocker:2003cw,Nandi:2002me}.

The second of the enumerated phases \textemdash~ which we
emphasize constitutes a new discovery presented in this essay 
 \textemdash~
does not contribute to
the redshift of the flavor oscillation clocks,
and it carries observable effects even for a single mass eigenstate.
It arises because the Lense-Thirring gravitational 
interaction energy 
associated with the
positive and negative helicity components
of $\mathcal{C}$  eigenspinors
has opposite signs, and hence induces opposite phases. Such an effect is 
entirely absent from eigenspinors of the 
 $\mathcal{Q}$ operator where both the right- and left- transforming
components carry the same 
Lense-Thirring gravitational 
interaction energy, and hence the same phases.

The former, i.e. the effect enumerated as
$\alpha$, is generic to both the eigenstates of the
$\mathcal{Q}$ and $\mathcal{C}$, while the latter,
 i.e. the effect arising from
item $\beta$,
is a unique characteristic of the 
$\mathcal{C}$ eigenstates.

As a result,
if the initial
eigenstate was associated with the spinor
$\lambda^S_{\{- +\}}(\p)$, then \textemdash~ if the test particle 
 is in the immediate vicinity of the surface \textemdash~ it oscillates
with frequency 
\beq
\Omega = \frac{G}{c^2}\left\vert
\frac{ \vec{\mathcal{J}}\cdot \widehat{\mathbf{h}} - 
3 \left(\vec{\mathcal{J}}
\cdot\widehat{\mathbf{r}}\right)
\left(\widehat{\mathbf{r}}\cdot  \widehat{\mathbf{h}}\right) }
{R^3}\right\vert\,,
\label{omega}\eeq
to an eigenstate associated with the spinor $\lambda^A_{\{-+\}}(\p)$. 
The
result (6) is valid in weak
field limit with involved speeds $v\ll c$, i.e., 
all the way to the vicinity of neutron stars.
In Eq. (6), $\widehat{\mathbf{h}}$ represents the direction to which helicity 
eigenstates in $\lambda(\p)$ refer to (and coincides with $\widehat{\mathbf{p}}$), 
$\widehat{\mathbf{r}}$ is
a unit position vector pointing from center of the idealized
astrophysical sphere to a general location on the sphere, and
the magnitude of $\vec{\mathcal{J}}$ is given by:
$
{\mathcal{J}} \approx \frac{2}{5} M R^2 \omega
$.  
If the state is considered stationary, i.e., at rest, on
the surface, then the probability of oscillation 
is $\sin^2(\Omega \tau)$. This expression readily generalizes 
to relativistic case, but then spatial variation of
$\Omega$ must be accounted for.

For accessing rough magnitude of the effect 
let us consider polar region of the gravitational source, 
with a test particle prepared
in such way that:
$\widehat{\mathbf{h}}=\widehat{\mathbf{r}} $. Then, we have  
\beq
\Omega = \left[\frac{4}{5}\left(\frac{G M}{c^2 R}\right)\right] 
\omega\,.\label{nr}
\eeq

The above results easily extend to a linear superposition of different 
mass eigenstates, specifically to neutrinos.
In that context (see Appendix for some technical details),
rotation in gravitational environments induces 
Majorana neutrinos to oscillate
from $\nu_e$ to $\overline{\nu}_e$.
In reference
to Eq. (7), note that 
\beq
 \frac{GM}{c^2 R} \approx\cases{ 
2.12\times 10^{-6},&  for Sun\cr
0.2,     &  {for Neutron stars\,}}
\eeq
while 
\beq
 \omega \approx\cases{ 
3\times 10^{-6}\; \mbox{s}^{-1},&    for Sun\cr
6.3\times10^{3}\; \mbox{s}^{-1},                &  
for  millisecond Neutron stars\,.}
\eeq
Therefore, should neutrinos be Majorana (for strong indications in that
direction, see below), the new effect is unobservable for solar neutrinos
with present experiments \cite{Eguchi:2003gg}. But the noted orders of 
astrophysical 
parameters indicate the effect may be significant in early universe 
and it carries importance for
supernovae explosions (where neutron stars are a significant
part of the gravitational environment). This
is so  because $\nu_e \rightleftharpoons
\overline{\nu}_e$ oscillations modify
energy transport. 

\textit{Concluding remarks.} \textemdash~ We end this essay with the 
observation 
that after decades of pioneering work, the Heidelberg-Moscow collaboration has,
in the last few years, presented first experimental evidence for the Majorana 
nature of $\nu_e $ and $\overline{\nu}_e$. The initial 3-$\sigma$ signal
now has better than 4-$\sigma$ significance
 \cite{Klapdor-Kleingrothaus:2001ke,Klapdor-Kleingrothaus:md,Klapdor-Kleingrothaus:gs,Klapdor-Kleingrothaus:2004ge}. 
Our essay shows that
Majorana neutrinos carry unsuspected sensitivity to gravitational environments
with rotational elements (which induce a new asymmetry). This asymmetry causes
$\nu_e \rightleftharpoons
\overline{\nu}_e$ oscillations with a frequency determined by the
gravitational environment. A neutrino sea composed entirely
of $\nu_e$ shall in time develop a 
$\overline{\nu}_e$ component inducing  a matter-antimatter asymmetry.
It is worth noting that
Raffelt \cite{Raffelt:1996ggr} has emphasized that 
core of a collapsing star, where matter density can reach as high as
 $3\times 10^{14}\;\mbox{g cm}^{-3}$ \textemdash~
with neutrino trapping density of about $10^{12}\;\mbox{g cm}^{-3}$
for $10\;\mbox{MeV}$ neutrinos \textemdash~is the only known 
astrophysical site, apart from early universe, where neutrinos
are in thermal equilibrium. It is here that 
the predicted $\nu_e \rightleftharpoons
\overline{\nu}_e$ oscillations may induce significant matter-antimatter
asymmetry while at the same time affecting the entire
evolution of the supernovae explosions.

\textit{Acknowledgments.} \textemdash~

\textit{This work was done, in part, under the auspices of the 
Mexican funding Agency, CONACyT, 
and was supported by its Project E-32067.
The author warmly thanks his wife, Dr. Irada Ahluwalia-Khalilova, for
reproducing the lost \LaTeX ~file from a printed copy.}

%.......................IMPORTANT ......................
% CQG -> Please do not edit out the paper titles.
%        The Letter has been written in such a manner
%        that the information contained in the titles
%        is essential for a better apprecaition of the
%        content.
%.......................................................

%\section*{References}

\appendix{
\textit{Appendix: A set of linearly connected eigenstates of} $\mathcal{C}$.} \textemdash~ 
There is a slight, 
and only apparent, asymmetry in the manner in which we introduced $\lambda(\p)$ spinors. 
A linearly related, and very useful, set of eigenstates of the $\mathcal{C}$ operator are:

\beq
\rho^S(\p) = 
\left(
\begin{array}{c}
\phi_R(\p)\\
-\sigma_2 \phi^\ast_R(\p)
\end{array}
\right)\,,\quad
\rho^A(\p) = 
\left(
\begin{array}{c}
\phi_R(\p)\\
\sigma_2 \phi^\ast_R(\p)
\end{array}
\right)\,
\eeq
with $
\mathcal{C}  \,\rho(\p) = \pm\, \rho(\p)
$, where the plus sign is for self conjugate spinors and minus sign is for antiself 
conjugate spinors. In the above equation $ \phi_R(\p) $ 
are massive right-handed Weyl spinors. Now, under 
Lorentz boosts $\sigma_2 \phi^\ast_R(\p)$ transform as left-handed spinors.

It can be shown that the sets  $\lambda(\p)$ and $\rho(\p)$ are not linearly independent.

The \textit{Majorana dual} is defined as:

\beq
\lambda^S(\p):\quad  
 \stackrel{\neg}{\lambda}^{S}_{\pm,\mp}\, (\p) \defn +\, [\rho^A_{\mp,\pm}\,(\p)]^\dagger   \gamma^0  
\eeq

\beq
\lambda^A(\p):\quad    
 \stackrel{\neg}{\lambda}^{A}_{\pm,\mp}\, (\p) \defn -\, [\rho^S_{\mp,\pm}\,(\p)]^\dagger   \gamma^0,  
\eeq
where
\beq
 \gamma^0 \defn \left(\begin{array}{rc} 0 & I\\ I & 0 \end{array}\right). 
\eeq
With this definition, the self/antiself conjugate  $\lambda(\p)$  spinors satisfy the following
orthonormality
\beq
 &&  \stackrel{\neg}{\lambda}^{S}_\eta (\p)\, \lambda^S_{\eta^\prime}(\p) = +\,   2m\, \delta_{\eta\eta^\prime}, \\
 &&  \stackrel{\neg}{\lambda}^{A}_\eta (\p) \,\lambda^A_{\eta^\prime}(\p) = -\,   2m \,\delta_{\eta\eta^\prime},   
\eeq
and completeness relations:
\beq 
\frac {1} {2m} \sum_\eta\left[\lambda^S_\eta(\p) \stackrel{\neg}{\lambda}^{S}_\eta(\p) - 
\lambda^A_\eta(\p)  \stackrel{\neg}{\lambda}^{A}_\eta(\p)\right] = I.
\eeq
In the above equations, the subscript $\eta$ ranges over two possibilities:$ \lbrace {+,-}\rbrace, \lbrace{-,+}\rbrace $.
The relations enumerated here help in obtaining results given in the main text. In particular, 
these relations allow for identification
 of the discussed oscillations with $ \nu_e\rightleftharpoons\overline{\nu}_e  $ oscillations.
Detailed analysis and interpretational issues shall be presented in Ref. \cite{ahluwalia:abr}


\begin{thebibliography}{100}

\bibitem{Lense:1918lte}
J.~Lense  and H.~Thirring, ``\"Uber den Einfluss der Eigenrotation 
der Zentralkörper auf die Bewegung der Planeten und Monde nach 
der Einsteinschen Gravitationstheorie,'' 
Phys.\ Z. {\bf 19} (1918) 156-63\\ 
English translation: B.~Mashhoon, F.~W.~Hehl, and D.~S.~Theiss, 
``On the gravitational effects of rotating masses: the 
Thirring-Lense papers,'' Gen.\ Rel.\ Grav. {\bf 16} (1984) 711


%\cite{Mukhopadhyay:2003ij}
\bibitem{Mukhopadhyay:2003ij}
B.~Mukhopadhyay and P.~Singh,
``Neutrino antineutrino asymmetry around rotating black holes,''
arXiv:gr-qc/0303053.
%%CITATION = GR-QC 0303053;%%


%\cite{Lambiase:2003fd}
\bibitem{Lambiase:2003fd}
G.~Lambiase and P.~Singh,
``Matter-antimatter asymmetry generated by loop quantum gravity,''
Phys.\ Lett.\ B {\bf 565} (2003) 27
[arXiv:gr-qc/0304051].
%%CITATION = GR-QC 0304051;%%


%\cite{Anandan:2003ev}
\bibitem{Anandan:2003ev}
J.~Anandan, N.~Dadhich and P.~Singh,
 ``Action based approach to the dynamics of extended bodies in General
Relativity,''
Int.\ J.\ Mod.\ Phys.\ D {\bf 12} (2003) 1651
[arXiv:gr-qc/0305063].
%%CITATION = GR-QC 0305063;%%


%\cite{Mohseni:2004br}
\bibitem{Mohseni:2004br}
M.~Mohseni,
``World-line deviation and spinning particles,''
arXiv:gr-qc/0403055.
%%CITATION = GR-QC 0403055;%%

%\cite{ahluwalia:abr}
\bibitem{ahluwalia:abr}
D.~V.~Ahluwalia-Khalilova, D.~Grumiller (in preparation).



%\cite{ramond:1989prb}
\bibitem{Ramond:1989prb}
P.~Ramond, \textit{Field Theory: A Modern Primer}
(Addison-Wesley Publishing Company, Redwood City, 1989).
%%%%CITATION


%\cite{Sanchez:2003rb}
\bibitem{Sanchez:2003rb}
M.~Sanchez {\it et al.}  [Soudan 2 Collaboration],
``Observation of atmospheric neutrino oscillations in Soudan 2,''
Phys.\ Rev.\ D {\bf 68} (2003) 113004
[arXiv:hep-ex/0307069].
%%CITATION = HEP-EX 0307069;%%





%\cite{Bellerive:2004dm}
\bibitem{Bellerive:2004dm}
A.~Bellerive  [SNO Collaboration],
``Constraints on neutrino mixing parameters with the SNO data,''
arXiv:hep-ex/0401018.
%%CITATION = HEP-EX 0401018;%%




%\cite{Kato:2003qx}
\bibitem{Kato:2003qx}
I.~Kato  [K2K Collaboration],
``The results of oscillation analysis in K2K experiment and an overview  of
JHF-nu experiment,''
arXiv:hep-ex/0306043.
%%CITATION = HEP-EX 0306043;%%




%\cite{Smy:2002rz}
\bibitem{Smy:2002rz}
M.~B.~Smy  [Super-Kamiokande collaboration],
 ``Solar neutrino precision measurements using all 1496 days of
Super-Kamiokande-I data,''
Nucl.\ Phys.\ Proc.\ Suppl.\  {\bf 118} (2003) 25
[arXiv:hep-ex/0208004].
%%CITATION = HEP-EX 0208004;%%



%\cite{Habig:2001ej}
\bibitem{Habig:2001ej}
A.~Habig  [Super-Kamiokande Collaboration],
``Discriminating between $\nu_\mu \leftrightarrow \nu_\tau$ 
and $\nu_\mu \leftrightarrow \nu_{sterile}$ in
atmospheric $\nu_\mu$ 
oscillations with the Super-Kamiokande detector,''
arXiv:hep-ex/0106025.
%%CITATION = HEP-EX 0106025;%%

%\cite{Ahluwalia:1996ev}
\bibitem{Ahluwalia:1996ev}
D.~V.~Ahluwalia and C.~Burgard,
``Gravitationally Induced Quantum Mechanical Phases 
and Neutrino Oscillations
in Astrophysical Environments,''
Gen.\ Rel.\ Grav.\  {\bf 28} (1996) 1161
[arXiv:gr-qc/9603008]; Erratum, Gen.\ Rel.\ Grav.\  {\bf 29} (1997) 681.
%%CITATION = GR-QC 9603008;%%


%\cite{Ahluwalia:1998jx}
\bibitem{Ahluwalia:1998jx}
D.~V.~Ahluwalia and C.~Burgard,
``Interplay of gravitation and linear superposition of different mass
eigenstates,''
Phys.\ Rev.\ D {\bf 57} (1998) 4724
[arXiv:gr-qc/9803013].


%%CITATION = GR-QC 9803013;%%
%\cite{Konno:kq}
\bibitem{Konno:kq}
K.~Konno and M.~Kasai,
``General Relativistic Effects Of Gravity In Quantum Mechanics: A Case  Of
Ultra-Relativistic, Spin 1/2 Particles,''
Prog.\ Theor.\ Phys.\  {\bf 100} (1998) 1145.
%%CITATION = PTPKA,100,1145;%%


%\cite{Wudka:2000rf}
\bibitem{Wudka:2000rf}
J.~Wudka,
``Mass dependence of the gravitationally-induced wave-function phase,''
Phys.\ Rev.\ D {\bf 64} (2001) 065009
[arXiv:gr-qc/0010077].
%%CITATION = GR-QC 0010077;%%

%\cite{Adak:2000tp}
\bibitem{Adak:2000tp}
M.~Adak, T.~Dereli and L.~H.~Ryder,
``Neutrino oscillations induced by space-time torsion,''
Class.\ Quant.\ Grav.\  {\bf 18} (2001) 1503
[arXiv:gr-qc/0103046].
%%CITATION = GR-QC 0103046;%%


%\cite{Crocker:2003cw}
\bibitem{Crocker:2003cw}
R.~M.~Crocker and D.~J.~Mortlock,
``Neutrino interferometry in curved spacetime,''
arXiv:hep-ph/0308168.
%%CITATION = HEP-PH 0308168;%%


%\cite{Nandi:2002me}
\bibitem{Nandi:2002me}
K.~K.~Nandi and Y.~Z.~Zhang,
 ``General relativistic effects on 
quantum interference and the principle of
equivalence,''
Phys.\ Rev.\ D {\bf 66} (2002) 063005
[arXiv:gr-qc/0208050].
%%CITATION = GR-QC 0208050;%%


%\cite{Eguchi:2003gg}
\bibitem{Eguchi:2003gg}
K.~Eguchi {\it et al.}  [KamLAND Collaboration],
``A high sensitivity search for $\overline{\nu}_e$'s from 
the sun and other sources at
KamLAND,''
Phys.\ Rev.\ Lett.\  {\bf 92} (2004) 071301
[arXiv:hep-ex/0310047].
%%CITATION = HEP-EX 0310047;%%


%\cite{Klapdor-Kleingrothaus:2001ke}
\bibitem{Klapdor-Kleingrothaus:2001ke}
H.~V.~Klapdor-Kleingrothaus, A.~Dietz, H.~L.~Harney and I.~V.~Krivosheina,
``Evidence for neutrinoless double beta decay,''
Mod.\ Phys.\ Lett.\ A {\bf 16} (2001) 2409
[arXiv:hep-ph/0201231].
%%CITATION = HEP-PH 0201231;%%


%\cite{Klapdor-Kleingrothaus:md}
\bibitem{Klapdor-Kleingrothaus:md}
H.~V.~Klapdor-Kleingrothaus, A.~Dietz and I.~V.~Krivosheina,
``Status Of Evidence For Neutrinoless Double Beta Decay,''
Found.\ Phys.\  {\bf 32} (2002) 1181
[Erratum-ibid.\  {\bf 33} (2003) 679]
[arXiv:hep-ph/0302248].
%%CITATION = HEP-PH 0302248;%%


%\cite{Klapdor-Kleingrothaus:gs}
\bibitem{Klapdor-Kleingrothaus:gs}
H.~V.~Klapdor-Kleingrothaus, A.~Dietz, 
I.~V.~Krivosheina, C.~Dorr and C.~Tomei [Heidelberg-Moscow Collaboration],
``Support Of Evidence For Neutrinoless Double Beta Decay,''
Phys.\ Lett.\ B {\bf 578} (2004) 54
[arXiv:hep-ph/0312171].
%%CITATION = HEP-PH 0312171;%%

%\cite{Klapdor-Kleingrothaus:2004ge}
\bibitem{Klapdor-Kleingrothaus:2004ge}
H.~V.~Klapdor-Kleingrothaus, A.~Dietz, I.~V.~Krivosheina and O.~Chkvorets,
 ``Data Acquisition and Analysis of the 76Ge Double Beta Experiment in Gran
Sasso 1990-2003,''
arXiv:hep-ph/0403018.
%%CITATION = HEP-PH 0403018;%%


\bibitem{Raffelt:1996ggr}
G.~G.~Raffelt,
\textit{Stars as laboratories for fundamental physics},
(University of Chicago Press, Chicago, 1996); see, ch. 11.

\end{thebibliography}
\end{document}